\documentclass{article}

\usepackage[resetlabels,labeled]{multibib}
\newcites{S}{Primary Sources Cites}

\usepackage{color}

\usepackage{xcolor,colortbl}
\xdefinecolor{gray95}{gray}{0.65}
\xdefinecolor{gray25}{gray}{0.8}

\usepackage{authblk}

\title{A Focused Mapping Study on Customization in Interactive Technologies for Autism}
\author[1]{Roberto E. Lopez-Herrejon}
\author[2]{Gerardo Herrera}
\author[2]{Javier Sevilla}
\affil[1]{Dept. Software Engineering and IT, \'{E}cole de technologie sup\'{e}rieure, University of Qu\'{e}bec, Canada. roberto.lopez@etsmtl.ca}
\affil[2]{University of Valencia, Spain. \{gerardo.herrera,javier.sevilla\}@uv.es}
\date{}                     
\begin{document}

\maketitle

\begin{abstract}
Autism Spectrum Disorder (ASD) is neurodevelopmental condition characterized by social interaction and communication difficulties, along with narrow and repetitive interests. Being an spectrum disorder, ASD affects individuals with a large range of combinations of challenges along dimensions such intelligence, social skills, or sensory processing. Hence, any interactive technology for ASD ought to be customizable to fit the particular profile of each individual that uses it. 
The goal of this paper is to characterize the support of customization in this area. To do so, we performed a focused study that identifies the dimensions of ASD where customization has been considered on wearable and natural surfaces technologies, two of the most promising  technologies for ASD, and assess the empirical evaluation that supports them. Our study revealed that, even though its critical importance, customization has fundamentally not been addressed in this domain and it opened avenues for research at the intersection of human-computer interaction and software engineering. 
  
\end{abstract}


\section{Introduction}
\label{sec:introduction}

Autism Spectrum Disorder (ASD) is neurodevelopmental condition characterized by social interaction and communication difficulties, along with narrow and repetitive interests. 
ASD affects individuals in multiple and combined ways along 
areas such intelligence, social skills (e.g. unable to interpret non-verbal cues), or sensorial processing (e.g. sensitivity to noise or lights)~\cite{ANYAS/ElKalioubyPB06}. In the autism community, a common saying is: \emph{"if you've met one person with autism you've met one person with autism"}\footnote{Quote authored by Stephen Shore, http://www.autismasperger.net/.}. This entails that individuals with autism have unique sets of challenges and needs that must be addressed to help their development and integration to society.
 
There is an extensive and long standing research on using computer-based systems, that spreads over more than four decades~\cite{Kientz2013}, whose driving goal is to support the needs of people with autism and their families.  Currently, digital libraries have hundreds of articles on the subject. This research has been summarized to certain degree in many literature reviews and surveys studies, e.g.~\cite{SAGE/GrynszpanWPG14,Hourcade2013,Kagohara2013,Stephenson2015,Bereznak2012,Wainer2011}.
However, and despite its critical importance there has not been a systematic study on how the proposed computer-based systems can be adapted or personalized to the particular needs of individuals with autism. From a Software Engineering perspective, this issue translates to the question: How are current computer-based technologies for autism customized? 

In this paper, we addressed this question by performing a mapping study that focuses on two of the most promising technologies for autism support namely, natural surfaces and wearables. 
We want to catalogue the personal characteristics that have been studied for these technologies, what type of support have they been used for, in what contexts they have been deployed, how they can be customized, and what empirical evidence supports them.
In contrast with standard systematic mapping studies, we select our primary sources from a set of already identified seminal articles on autism and interactive technologies that contains over 400 articles~\cite{Kientz2013}. 
We also complement our primary sources selection with additional searches particular to personalization and customization topics.

Our focused mapping study indeed corroborated the lack of research on customization for technologies that support autism. 
Based on this fact, we draw connections to \textit{Software Product Line Engineering (SPLE)} an area of research and practice of Software Engineering whose goal is to support mass costumization of software systems (e.g.~\cite{SPLE}). Extensive research and practice of SPLE spanning over more than two decades attest the benefits of applying SPLE principles.
We argue that such connections between SPLE and interactive technologies for autism open up avenues for further research that will ultimately help to address the need of customized computer-based solutions for persons with autism.


\nociteS{S1-IWVR/ParesCDFFGS04}
\nociteS{S2-PUCom/SitdhisanguanCDO12}
\nociteS{S3-PUCom/Keay-BrightW12}
\nociteS{S4-EMBS/AmirabdollahianRDJ11}
\nociteS{S5-Autism/FarrYR2010}
\nociteS{S6-IDC/FarrYHH10}
\nociteS{S7-FIE/DrainRLR11}
\nociteS{S8-HFCS/HailpernKH09}
\nociteS{S9-DISC/HailpernHKBK12}
\nociteS{S10-TNSRE/PioggiaIFAMD05}
\nociteS{S11-SIA/BlocherP02}
\nociteS{S12-StricklandMCO07}
\nociteS{S13-RosenbloomMWM16}
\nociteS{S14-MechlingAFB15}

\section{Background}
\label{sec:background}

In this section we provide a basic background knowledge on autism and describe the core ideas of software customization within the development paradigm of Software Product Line Engineering.

\subsection{Autism Basics}
\label{subsec:autism}

Autism Spectrum Disorder (ASD) is a neurodevelopmental disorder characterized by impaired social interaction and communication, and restricted and repetitive behavior~\cite{APA}. ASD is diagnosed in at least 1\% of the population, and diagnoses are more common amongst males than females~\cite{Baird2006}. Autism can have profound impact upon learning and it is estimated that 54\% of individuals with autism also have intellectual disability/learning disability (Center for Disease Control CDC\footnote{Center for Disease Control (CDC) https://www.cdc.gov/ncbddd/autism/data.html}). 

Today, no medical treatment is available for the core symptoms of autism. Early intervention programs, usually aimed at children from 0 to 6 years old,  have been demonstrated effective for supporting the development of a relevant percentage of children with autism. 
The most effective programs have a behavioural base (e.g. Applied Behaviour Analysis (ABA)~\cite{Cohen2006}) or cognitive-behavioural base (e.g. Early Start Denver Model (ESDM)~\cite{Rogers2010}). 
Within this later program, for example, there is a developmental curriculum (the ESDM checklist) that is highly personalised for each child with autism and intervention objectives are redefined every three months in order to adapt to the child progress. 

Intervention programs in autism can be classified as focused-intervention (FI) programs~\cite{Wong2015}, usually designed for improving a particular ability (or a reduced set of abilities) or comprehensive treatment models (CTM) that are much wider and are based on a holistic approach of the child development~\cite{Odom2010}. Some of these programs use technologies as a basis for documenting the child progress, and some other use technology for very particular tasks. However, none of these programs are genuinely based on any particular technology. When available, innovative technologies are used for Focused Interventions rather than as Comprehensive Treatment Models. Most research evidence available on technologies for ASD rely mainly on the use of particular communicator apps on tablet devices while the evidence on other areas seems to be anecdotal or at least not enough explored~\cite{Lorah2014}.

%

\subsection{Software Customization}

Software customization is the process of developing software for a specific user or purpose. This is in contrast with Commercial off-the-shelf (COTS) software products that are developed targeting very general market segment.

\textit{Software Product Lines (SPLs)} are families of related systems whose members are distinguished by the set of features they provide, where a \textit{feature} is an increment in functionality~\cite{DBLP:journals/tse/BatorySR04,SPLE}. 
\textit{Software Product Line Engineering (SPLE)} refers to the paradigm of developing SPLs. Typical SPLE efforts involve a large number of features that are combined in complex feature relations yielding a large number of individual software systems that must be effectively and efficiently designed, implemented and managed. 
Hence SPLs' main goal is to enable mass customization of software products.
There is an extensive body of research over more than two decades that attests to the benefits of SPL practices and that has proposed multiple SPLE approaches, methods, and techniques~\cite{SPLE,DBLP:journals/infsof/HeradioPFCH16, DBLP:journals/tse/GalsterWTMA14,DBLP:journals/infsof/ChenB11,ProductLinesInAction,DBLP:book/SPLEngManIssues}.

A key concept in SPLE is \textit{variability} which is the capacity of software artifacts to vary.  The effective management and realization of variability lie at the core of successful SPL development~\cite{DBLP:journals/spe/SvahnbergGB05}. This capacity to vary or change is to reflect the different possible combinations of features required for each individual software system.
Several forms of \textit{variability models} have been proposed that succinctly and formally express all the desired combination of features~\cite{KCH+90,DBLP:journals/is/BenavidesSC10}. Similarly, several variability implementation mechanisms have been studied~\cite{KCH+90,FOSD-book}.

The most common applications of SPLs have been in the automotive industry, embedded systems, avionics, and medical equipment\footnote{Software Product Line Conference Hall of FAME, http://splc.net/fame.html}. To the best of our knowledge, the domain of Health Informatics remains largely unexplored. 
It is widely accepted that there is no "best technology for autism" and different efforts are being made to facilitate the way of finding the best technology for each individual with autism.
Hence, given the diverse nature of this condition across individuals, we believe SPL practices could be exploited for better customizing the technological support needed.

\section{Focused Mapping Study}
\label{sec:study}

A mapping study is a secondary study intended to identify and classify the set of publications on a topic~\cite{book/KitchenhamBB15evidence}. We performed a mapping study on customization support for technologies for autism along the standard guidelines provided by Petersen, Kitcheman, and others~\cite{conf/ease/Petersen08,book/KitchenhamBB15evidence}. However because of the extensive body of work available on technologies for autism, we decided to focus on two concrete technologies and to extract the primary sources from an existing repository of selected works which we extended with searches to consider customization terms. Next we describe the process we carried out, while in Section~\ref{sec:results-analysis} we present the results we obtained and their analysis.

\subsection{Research questions}
\label{subsec:questions}

Recall that the main goal of our focused study is to provide an overview of the support of customization in interactive technologies for autism. 
We chose two concrete technologies~\cite{Kientz2013}: \textit{i)} \textit{natural interfaces} that use input devices beyond traditional keyword and mice, for example pens, speech, gesture, eye-tracking, etc., and \textit{ii)} \textit{wearables} which include sensors (e.g. hear rate, microphones, etc.) both in the environment and the body to collect data and input. We studied these two types of technologies as they hold the most promise for providing support across many needs of people with autism (e.g.~\cite{WearablesPicard16}).

We present now the research questions considered by our study to achieve this goal.

\begin{itemize}

\item \textbf{\textit{RQ1. What personal characteristics of people with autism have been considered in natural interfaces and wearable technologies?}} 

\textit{Rationale:} People with autism have differences in sensorial abilities (e.g. sound, vision, touch, etc.), intellectual abilities, interaction capabilities with computer-based technologies, etc. This question aims to catalogue the personal dimensions that have been catered for with these technologies

\item \textbf{\textit{RQ2. What are the developmental domains that have been addressed by natural interfaces and wearable technologies?}} 

\textit{Rationale:} Interactive technologies for autism are conceived to meet specific needs and improve concrete challenges in certain developmental domains such as language and communication skills, social interactions, etc. The purpose of this question is to catalogue the domains that have been explored using the selected technologies. 

\item \textbf{\textit{RQ3. What are the context where natural user interfaces and wearable technologies have been deployed?}} 

\textit{Rationale:} This question aims to catalogue the settings, e.g. home or school, where these technologies have been used for autism.

\item \textbf{\textit{RQ4. What is the empirical support of the proposed approaches?}}

\textit{Rationale:} This question aims to collect the empirical evidence that supports each approach identified. We are interested in the type of experimentation employed, the number of participants, and the availability of the technology for replication (i.e. access to software, hardware description, etc.)

\item \textbf{\textit{RQ5. What customization support is provided by the proposed approaches?}}

\textit{Rationale:} This question aims to cataloque customization support across the dimensions of personal characteristics (question RQ1) and developmental domains (question RQ2).

\end{itemize}

\subsection{Conduct Search for Primary Sources}
\label{subsec:primary-sources}

In this section we described the process followed to obtain our primary sources. 
As a first step, we selected seminal papers on the two technologies we focused on, natural interfaces and wearables, from the repository  associated with the book Interactive Technologies for Autism\footnote{Located in Mendeley social network in a group with the same name as the book, https://www.mendeley.com} written by Kientz, Goodwin, Hayes, and Abowd who are recognized experts in the area~\cite{Kientz2013}. 

As a second step, we performed a search using Web of Science\footnote{http://apps.webofknowledge.com/} with the goal of retrieving articles on customization of any interactive technologies for autism, not only on our two selected ones. We employed this search engine because its advanced query capabilities and because it indexes all the publication outlets on autism and technology. 
We used terms commonly employed in the area (e.g. personalization, customization, ASD) that we apply in the following query\footnote{TS=Topic Search, ASC=Autism Spectrum Condition}:

\begin{center}
\begin{tabular}{p{10cm}}
\texttt{
(TS=((Autis* OR ASD OR ASC OR "Asperger Syndrome" OR "Pervasive Developmental Disorder" OR PDD*) AND (Technolog* OR Computer* OR
Virtual* OR Robot*) AND (Custom* or Personali*))) AND (Search Language = English) AND (Document type= Article) AND (Timespan= 2000-2016) AND (Indexes: SCI-EXPANDED, SSCI, A\&HCI, CPCI-S, CPCI-SSH, ESCI, CCR-EXPANDED, IC.)
}
\end{tabular}
\end{center}

\subsection{Inclusion and Exclusion}
\label{subsec:inclusion-exclusion}

The basic criterion for inclusion in our study was a clear application of a computer-based technology for supporting a therapy or intervention in relation to autism, where individuals with autism participated in the design, validation or evaluation of the technology.

The criteria to exclude papers in our study was:
\textit{i)} papers which did not describe a technology that supports any intervention or therapy (e.g. paper that describes biosignal monitoring tool), 
\textit{ii)} individuals with autism were not involved at any stage of design, validation or evaluation of the technology, 
\textit{iii)}
papers not written in English, \textit{iv)} vision or position papers that had no implementation to back them up, \textit{v)} graduate or undergraduate dissertations and thesis, and \textit{vi)} non peer-reviewed documents such as technical reports.

During the screening process we looked for the search terms in the title, abstract and keywords and whenever necessary at the introduction or at other places of the paper. 
The decision on whether or not to include a paper was most of the times straightforward, in other words, a clear application of computer-based technologies to autism with the participation of individuals with the condition was easily drawn.

%
%

\subsection{Classification}
\label{subsec:classification}

In contrast with standard mapping studies with classification terms emerge from the analysis of primary sources, we use instead common classification terminology and schemes within the area of interactive technologies for autism. 
Basically, we classified the primary sources along different dimensions in each question except for the last question on customization support that cuts across several dimensions. Next we outline these classification dimensions and provide a rationale for their selection.

\subsubsection{Dimensions for personal characteristics} Here we use four dimensions to cover important aspects of the personal profile of persons with autism which can be the subject to customization~\cite{Kientz2013,Robinson2011}.

\begin{itemize}

\item \emph{Sensorial} dimension where we consider each sense (e.g. vision, hearing, touch, etc.) a category.

\item \emph{Intellectual} ability to consider the different conditions of intellectual, cognitive or learning disability.

\item \emph{Interaction} capability that considers what forms of human-computer interactions are customized. Some examples, are pushing screen buttons, sliding fingers in tablet surfaces, or recognizing speech commands.

\item \emph{Level of support needed to use the technology} makes referece to the degree of support a particular technology needs to be used. It goes from \emph{none} support, to \emph{medium support} when for example basic training is needed, to \emph{high support} where continuous support or complex training is needed to use a particular technology. 

\end{itemize}

\subsubsection{Developmental domains} For this question we adhere to the classification scheme put forward by Kientz et al.~\cite{Kientz2013}. We use this scheme because it is a common reference framework for researchers working on the field of interactive autism technologies. We summarize the domains as follows:

\begin{itemize}

\item \emph{Social and emotional} covers skills that focus on emotion recognition and support behaviours that improve social interactions.

\item \emph{Language and communication} covers skills that aim at improving linguistics aspects such as prosody and syntax, as well as language acquisition and reading skills.  

\item \emph{Restrictive and repetitive behaviours} covers addressing these types of behaviours characteristic of autism that manifest at different cognitive levels.

\item \emph{Academic} that refers to skills commonly associated to formal education,  such literacy, numeracy, etc.

\item \emph{Life and vocational} that refers to skills related to daily life (e.g. grocery shopping) or to work scenarios.

\item \emph{Sensory, physiological responding and motor} that refer to the individual's profile in relation to issues such as sensory regulation, perception, or fine and gross motor movements.

\end{itemize}

\subsubsection{Physical Context settings}
For this question we consider the context setting where our selected technologies have been deployed. Here we also follow the work by Kientz et al.~\cite{Kientz2013}, who classify the context in: \emph{home}, \emph{school}, \emph{research lab} (i.e. at a research institution), \emph{clinic} (e.g. doctor's or therapists office), and \emph{community} (e.g. parks, stores, etc.).


\subsubsection{Empirical support}
Research design, or experimental design, is an important knowledge area with a wide range of alternatives for approaching each particular study. For the scope of this review, we have considered only the two principal types of design that are applied in most technological studies~\cite{Kientz2013,WohlinExperimentalSE12}. These are single subject research design and group research design. Single subject research design refers to research in which the subject serves as his/her own control, rather than using another individual/group. Horner et al. developed guidelines and quality indicators for interventions aimed at students with special educational needs~\cite{Horner2005}. Group research design (where two-group posttest-only randomized experiments are the simplest type) refers to research where one group of participants (treatment group) is compared to another group (control group) with participants in both groups balanced around variables such as age, IQ or severity of autism symptoms around social communication or restrictive/repetitive behaviours. Gersten et al. developed quality indicators for group research designs~\cite{Gersten2005}.

We classify also the availability of the resources for replication, for instance open sources and documentation. For this latter category, we use \emph{none}, \emph{partial}, and \emph{full} depending on the degree of availability.


\subsection{Data Extraction and Mapping Study}
\label{subsec:data}

For gathering the data we proceeded with the following steps: 

\begin{enumerate}

\item We created an spreadsheet to collect the classification information. The spreadsheet contained the following data fields: 
\textit{i)} sensorial characteristics,  \textit{ii)} intellectual disability characteristics, \textit{iii)} interaction capabilities, \textit{iv)} level of support needed, \textit{v)} developmental domains, \textit{vi)} physical context settings, \textit{vii)} number of participants involved, \textit{viii)} type of empirical evaluation,  \textit{ix)} availability of resources for replication, and \textit{x)} forms of customization supported.

\item We formed two groups to carry out the classification task independently.

\item We held a meeting to pilot the classification terms.
In this meeting each group presented its classification of a group of five selected primary sources. Any discrepancies were discussed and analyzed to homogenize the classification criteria.

\item The two teams performed the classification of all primary sources independently. 

\item We held a second meeting where the classification for every single paper for each criterion was discussed until a consensus was reached. 

\end{enumerate}

The effort to gather the data varied between papers but overall it was a simple task to find all the classification information required. The most time-consuming part was in some cases finding out the empirical and customization support information.

\section{Results and Analysis}
\label{sec:results-analysis}

In this section we first describe the primary sources that were identified, followed by the results obtained for each research question and the threats to validity that relate to our study.


\begin{table*}[t]
\center
\caption{Technology, interaction forms, and customization support summary}
\footnotesize
\begin{tabular}{p{1.0cm}|p{3.0cm}|p{3.0cm}|p{3.5cm}}
\hline
 \textbf{Primary source} & \textbf{Technology} & \textbf{Interaction forms} & \textbf{Customization support} \\ \hline 
 S1 & retroprojection, computer vision, natural surfaces & body interaction
& interaction demands adaptive to child behaviour\\ \hline
 S2 & tangible user interfaces &  object manipulation
 &  no support provided \\ \hline
 S3 & multi-touch screen & touch screen &  no support provided \\ \hline 
 S4 & robot & robot interaction &  customizable touch measurement thresholds \\ \hline
 S5 & tangible user interface &  object manipulation &  no support provided \\ \hline  
 S6 & tangible user interface &  object manipulation & customizable interaction contents \\ \hline  
 S7 & modular electronic system &  object manipulation & one open-ended customizable activity from a set of tasks \\ \hline  
 S8 &  speech recognition and visual feedback &  speech commands & no support provided \\ \hline  
 S9 & speech recognition and visual feedback  &  monitors and microphones & no support provided \\ \hline
 S10 & robot &  robot interaction &  no support provided \\ \hline  
 S11 & tangible user interface &  object manipulation &  multiple customization dimensions: multimedia I/O, emotions set, user interface, reinforcement  \\ \hline  
 S12 & immersive virtual reality &  headset, motion tracker, 3D joystick, mouse, keyboard
&  multiple customization dimensions: scene complexity, instructional guides, sounds, forms of control\\ \hline  
 S13 & smartphone &  screen interaction & customization for prompts, recording, data monitoring \\ \hline  
 S14 & video recording and playing &  video watching & highlights importance of custom-made videos  \\ \hline             

 \hline 
 \end{tabular}
\label{tab:tech-int-comp}
\end{table*}

\subsection{Primary sources selection}
\label{subsec:selection}

As described in Section~\ref{subsec:customization-support} our first selection step was from a repository of seminal literature in the area~\cite{Kientz2013}. This repository contains, at the time of writing, over 400 bibliography references. As mentioned before, in their book Kientz et al. propose a classification taxonomy  for mapping the available research literature. Unfortunately, this taxonomy was not applied to all the articles in their repository. Instead, it is only illustrated for a small sample. Nonetheless, their book is structured such that each chapter corresponds to a type of technology.
Hence following the chapter classification we extracted the references that corresponded to the chapters on our two focused technologies, natural user  intefaces (chapter 10 in~\cite{Kientz2013}) and wearables (chapter 8 in~\cite{Kientz2013}). This step resulted in 39 primary sources identified.

In our second step we carried out the query research which yielded 15 primary sources, of which one as already identified in the first step. At this stage we proceeded with a more in depth review applying the inclusion and exclusion criteria to the 53 unique sources identified. We soon realized that many of them needed to be excluded, mainly because they did not support a therapy or intervention approach,  were targeted to parents or therapists, or described technologies for monitoring purposes. The latter was the case for the articles in the wearable technologies.

After exclusion process we finalized with a list of 14 primary sources listed in the Primary Sources Cites section.

\subsection{Personal characteristics}
\label{sec:personal}

In this section we summarize our findings regarding the fours aspects of the personal characteristics of persons with autism that we considered for our study.

\subsubsection{Sensorial dimensions}
\label{subsub:sensorial}

Table~\ref{tab:sensorial} presents a summary of our findings. It is a well-known fact that people with autism favours visual information for many daily routine and learning tasks. So, it is not a surprise that except one primary study all of them exploit this sense. The only exception was the work of Amirabdollahian et al. that fundamentally focuses on touch~\citeS{S4-EMBS/AmirabdollahianRDJ11}, whereby the children participating in the therapy can only interact with a robot by touching it. The second most frequent sense was hearing with 10 primary sources, with a wide range of options from watching videos in the work of Mechling et al.~\citeS{S14-MechlingAFB15}, to several forms of speech recognition and synthesis~\citeS{S8-HFCS/HailpernKH09,S9-DISC/HailpernHKBK12}. The third place was touch with 9 primary sources. Here the most common form was using multi-touch screens and smartphones~\citeS{S3-PUCom/Keay-BrightW12,S13-RosenbloomMWM16}. The fourth place is the propioception sense with 3 primary sources. This sense refers to capability of inferring the relative position of the parts of the body and strength needed to carry out movement. Here, for instance, children needed to move objects such as lego-like bricks, e.g.~\citeS{S5-Autism/FarrYR2010}. The last sense, vestibular perception, refers to the capacity of balance and orientation in space. In our case it was a virtual reality system for fire safety training~\citeS{S12-StricklandMCO07}.

\begin{table}[h!]
\center
\caption{Sensorial dimensions summary}
\footnotesize
\begin{tabular}{p{1.8cm}cp{8.0cm}}
\hline
 \textbf{Sense} & No & \textbf{Primary Sources Identifiers} \\ \hline 

Vision   & 13 & S1, S2, S3, S5, S6, S7, S8, S9, S10, S11, S12, S13, S14 \\ 
Hearing & 10 & S1, S3, S6, S8, S9, S10, S11, S12, S13, S14 \\
Touch & 9 & S2, S3, S4, S5, S6, S7, S10, S11, S13 \\
Propioception & 3 & S5, S7, S12 \\
Vestibular perception & 1 & S12 \\
 \hline 
 \end{tabular}
\label{tab:sensorial}
\end{table}

\subsubsection{Intellectual Disability}
\label{subsub:intelectual}

Table~\ref{tab:intelectual} summarizes our findings for intellectual disability. Within the autism community, intellectual disability is broadly categorized  depending on the degree it affects the individual, ranging from \emph{none} to \emph{severe}. The approaches found in our study were more or less distributed across the spectrum, with \emph{severe} being the most frequent one followed by \emph{none}, \emph{moderate}, and \emph{mild}. For three primary sources it was not possible to determine the level(s) of disability they considered.

~\newline

\begin{table}[h!]
\center
\caption{Intellectual disability summary}
\footnotesize
\begin{tabular}{p{1.8cm}cp{5.0cm}}
\hline
 \textbf{Level} & No & \textbf{Primary Sources Identifiers} \\ \hline 
Severe & 6 & S1, S2, S3, S6, S8, S11 \\
 
None & 5 & S5, S7, S10, S11, S13 \\
Moderate & 4 & S6, S11, S12, S14 \\
Mild & 3 & S5, S7, S11 \\
Not possible to determine & 3 & S4, S9, S12  \\
 \hline 
 \end{tabular}
\label{tab:intelectual}
\end{table}

\subsubsection{Interaction Capability}
\label{subsub:interaction}

Table~\ref{tab:tech-int-comp} summarizes the different interactions forms found by our study. The most common form was manipulation of objects with 5 primary sources (S2, S5, S6, S7, S10) where children interacted by moving items, for instance lego-like bricks~\citeS{S5-Autism/FarrYR2010} or toys in a castle setting~\citeS{S6-IDC/FarrYHH10}.
Interaction with robots involved touching, looking and talking to them and was used primarily for emotion recognition and skills~\citeS{S4-EMBS/AmirabdollahianRDJ11,S10-TNSRE/PioggiaIFAMD05}.
Speech involved visualizing elements of spoken words~\citeS{S8-HFCS/HailpernKH09,S9-DISC/HailpernHKBK12}.
The rest of primary sources used varied forms of interactions: full body movement~\citeS{S1-IWVR/ParesCDFFGS04} to virtual reality~\citeS{S12-StricklandMCO07}, different combinations of multimedia interactions with screens~\citeS{S3-PUCom/Keay-BrightW12} and videos~\citeS{S14-MechlingAFB15}, or using mobile phones~\citeS{S13-RosenbloomMWM16}. 


%
%
%
%

\subsubsection{Level of Support Needed}
\label{subsub:support}

Table~\ref{tab:support} summarizes our findings for the level of support needed by the people with autism to interact and use the systems. Most of our primary sources, 11 instances, do not require support for their use. The 3 primary sources that do, they require basic training for instance to learn how to use an app~\citeS{S13-RosenbloomMWM16}. For the work of Mechling et al.~\citeS{S14-MechlingAFB15}, even though it involved recording and watching videos with general purpose tools, it was not possible to determine the degree to which people with autism interacted with such tools.


\begin{table}[h!]
\center
\caption{Level of support summary}
\footnotesize
\begin{tabular}{p{1.8cm}cp{7.0cm}}
\hline
 \textbf{Level} & No & \textbf{Primary Sources Identifiers} \\ \hline 
 
No support & 11 & S1, S2, S3, S4, S5, S6, S7, S8, S9, S10, S11 \\
Medium & 3 & S7, S12, S13 \\
Not possible to determine & 1 & S14 \\
 \hline 
 \end{tabular}
\label{tab:support}
\end{table}


\subsection{Developmental domains}
\label{sec:developmental}



Table~\ref{tab:developmental} summarizes the developmental domains. We found that our primary sources covered all the domains proposed by Kientz et al.~\cite{Kientz2013}. However, in addition we identified an extra category \emph{leisure} where an important purpose of the systems was that people with autism have fun an enjoy themselves. The most frequent category was "social and emotional" with 5 primary sources. For example, in the work by Blocher and Picard, they employ soft toys with different expressions to help children recognize emotions~\citeS{S11-SIA/BlocherP02}.
The second most frequent category was leisure with 4 primary studies. For instance, the MEDIATE project created an interactive room where children could freely play with different sources of light, sound, and surfaces~\citeS{S1-IWVR/ParesCDFFGS04}.
The next three categories have 3 primary sources each. An example in the academic category was presented by the work of Rosenbloom et al. who developed an app for self-monitoring behaviour for inclusion of children in conventional primary school~\citeS{S13-RosenbloomMWM16}. As another example, the work by Farr et al. provided an interactive castle setting to measure and monitor repetitive behaviour patterns in a game context~\citeS{S6-IDC/FarrYHH10}. The work by Amirabdollahian et al. employed touch as a form of sensory interaction with a robotic face~\citeS{S4-EMBS/AmirabdollahianRDJ11}. In the realm of lanauge and communication, the work of Halpern and colleages focus on visualizing different patterns in words to help children improve their speech skills~\citeS{S8-HFCS/HailpernKH09,S9-DISC/HailpernHKBK12}. Lastly, the work by Strickland et al. focused on life skills such as fire safety~\citeS{S12-StricklandMCO07}, whereas the work by Mechling et al. address like skills such as loading a dishwasher~\citeS{S14-MechlingAFB15}.


\begin{table}[h!]
\center
\caption{Developmental domains summary}
\footnotesize
\begin{tabular}{p{6.0cm}cp{4.5cm}}
\hline
 \textbf{Domain} & No & \textbf{Primary Sources Identifiers} \\ \hline 
 
Social and Emotional & 6 & S3, S5, S6, S7, S10, S11 \\
Leisure & 4 & S1, S3, S5, S6 \\
Academic & 3 & S2, S7, S13 \\
Restrictive and repetitive behaviours & 3 & S1, S5, S6 \\
Sensory, physiological responding and motor & 3 & S1, S3, S4 \\
Language and communication & 2 & S8, S9 \\
Life and vocational & 2 & S12, S14 \\
 \hline 
 \end{tabular}
\label{tab:developmental}
\end{table}


\subsection{Physical context settings}
\label{sec:context}


Table~\ref{tab:context} summarizes our findings for context settings. We found that schools and research laboratories are respectively the most predominant settings with 8 and 6 primary sources. An example from school setting is the work by Drain et al. who introduce lego-like electronic modules to stimulate social interaction~\citeS{S7-FIE/DrainRLR11}. An example from a research lab setting is the project MEDIATE where the interactive room needed to be located in the laboratory because of its technical complexity~\citeS{S1-IWVR/ParesCDFFGS04}.  
On third place of frequency, the clinical setting had 2 primary sources. An advantage of this setting was that it does not disrupt the therapy schedule of the participating children~\citeS{S2-PUCom/SitdhisanguanCDO12}.
Only one primary source was found for a home setting, where Strickland and colleagues teach fire safety~\citeS{S12-StricklandMCO07}.
Lastly, for one primary source we were not able to determine the setting where it was actually applied~\citeS{S6-IDC/FarrYHH10}.


\begin{table}[h!]
\center
\caption{Context settings summary}
\footnotesize
\begin{tabular}{p{2.3cm}cp{5.5cm}}
\hline
 \textbf{Context} & No & \textbf{Primary Sources Identifiers} \\ \hline 
 
 School & 8 & S3, S4, S5, S7, S9, S12, S13, S14 \\
 Research lab & 6 & S1, S8, S9, S10, S11, S12 \\
 Clinic & 2 & S2, S3 \\
 Home & 1 & S12 \\
 Not possible to determine & 1 & S6 \\
 \hline 
 \end{tabular}
\label{tab:context}
\end{table}


\subsection{Empirical support}
\label{sec:empirical}



Table~\ref{tab:empirical} summarizes our findings for empirical support. This table shows that single subject was by far the most prevalent type of research design with 10 primary sources, distantly followed by group research design with 2 primary sources. Additionally, for 2 primary sources for which it was not possible to determine what type of empirical evaluation they performed (see ~\citeS{S1-IWVR/ParesCDFFGS04,S4-EMBS/AmirabdollahianRDJ11}).

On the ten single subject primary sources, the number of participants ranged from one (see ~\citeS{S13-RosenbloomMWM16}) to 16. Six primary sources provided partial information for replication, most frequently either the research design details were missing or the details of the hardware or software were not publicly available. Three primary sources provided no information for replication. Only the work by Hailpern et al. provided full details for replication~\citeS{S8-HFCS/HailpernKH09}.

Of the group research design, the work by Sitdhisanguan et al. on tangible user interfaces employed 32 participants but did not provide details for replication~\citeS{S2-PUCom/SitdhisanguanCDO12}. In contrast, the work by Farr et al. included 12 participants and provided full details for replication~\citeS{S5-Autism/FarrYR2010}. 

Despite the critical importance of empirical support for any interactive technology, only two primary sources provided enough details to enable replication. Though encouraging, we argue this is an aspect that needs improvement across all the different research communities that work on the area.

\begin{table}
\center
\caption{Empirical support summary}
\footnotesize
\begin{tabular}{p{4.0cm}cp{2.5cm}}
\hline

\cellcolor{gray25} Single subject design & \cellcolor{gray25} No & \cellcolor{gray25} Level \\ \hline

S12 &  16 & None \\
S3	&  13 &  Partial \\
S6 & 12 & None \\	
S7	& 6	&	Partial \\
S11	& 6	& None	\\
S8	& 5	&	Full \\
S14	& 4	& Partial \\
S9 & 2 & Partial \\
S10	& 2	& Partial \\
S13	& 1	& Partial \\

\hline 
\cellcolor{gray25} Group design & \cellcolor{gray25} No & \cellcolor{gray25} Level \\ \hline
S2 &	32	& None \\
S5 & 12 & 	Full \\		


\hline 
\cellcolor{gray25} Not possible to determine & \cellcolor{gray25} No & \cellcolor{gray25} Level \\ \hline
S1 & 11 & Partial \\ 
S4 & NA & None \\

 \hline 
 \end{tabular}
\label{tab:empirical}
\end{table}



\subsection{Customization support}
\label{subsec:customization-support}

In this section we describe for each primary source what forms of customization were considered across the dimensions of our study. Last column of Table~\ref{tab:tech-int-comp} summarizes our findings. We present the description from higher to lower level of support.

Work by Blocher and Picard describes a system to teach children to recognize emotions~\citeS{S11-SIA/BlocherP02}. This system has customization as one of its core tenets. The therapists can customize the videos and audio used for input and output, the set of emotions to elicit, the user interface, and the types of feedback.
Strickland et al. provide customization for a virtual reality system for street and fire safety training that supports customization for scene complexity (e.g. car and number of colors), instructional guides (e.g. figures of dog, rabbit, etc.), sounds, and forms of control~\citeS{S12-StricklandMCO07}.
The work by Farr et at. allows children to customize the contents of interactions of several elements in a knight castle set~\citeS{S6-IDC/FarrYHH10}. For example, children can record voice messages that are played when a character is placed on certain locations of the castle.

~\newline
Work by Rosenbloom et al. developed an application that can be customized for prompts (e.g. chime or flash, frequency, etc.), recording, and data monitoring~\citeS{S13-RosenbloomMWM16}.
In the MEDIATE project customization is supported in their interactive room by providing capability to recognize behaviors and adapt the interactions accordingly to each child, i.e. raising or lowering the interaction demands~\citeS{S1-IWVR/ParesCDFFGS04}.
The work by Amirabdollahian et al. allows for customization of touch threshold measurements in child-robot interactions, in other words the minimum and maximum measures are ajusted to each child using their robot~\citeS{S4-EMBS/AmirabdollahianRDJ11}.

The work by Sitdhisanguan et al. do not provide support for customization because they want to assess learning outcomes of a computer-based training system~\citeS{S2-PUCom/SitdhisanguanCDO12}. However, they present valuable design guidelines that could be used to provide customization, for example the sizes of the table tops and object used, different use of sound, options to provide feedback on correct and incorrect answers, or choice of color for foreground and background contrast.
Drain et al use a lego-like platform called eBlocks for activities aim to improvie learning and social skills~\citeS{S7-FIE/DrainRLR11}. Their work sets up a set of tasks to accomplished, but has one open-ended (i.e. customizable) activity where students can build a small system of their own choosing.
Hailper et al. do not provide support for customization for their visual feedback on vocalization approach; however, they do acknowledge the potential benefit it can have~\citeS{S8-HFCS/HailpernKH09}. Their subsequent work also highlights the interest and potential benefits of customization but no provision is made to provide it~\citeS{S9-DISC/HailpernHKBK12}.
The work of Mechling et al. simply highlights the significant impact that custom-made videos have for learning daily living tasks~\citeS{S14-MechlingAFB15}.
The work by Keay-Bright et al. do not make provisions for customization~\citeS{S3-PUCom/Keay-BrightW12}. Similar situation is the work by Farr et al~\citeS{S5-Autism/FarrYR2010}, and Pioggia et al.~\citeS{S10-TNSRE/PioggiaIFAMD05}.

Our study confirmed that even though the importance of customizing technologies for autism is widely accepted and acknowledged, it is still not thoroughly supported. This is in particular more important when dealing with the senses of vision, hearing, and touch that have been shown to have the  largest abnormalties in people with autism~\cite{Leekam2007}.

\subsection{Threats to validity}
\label{subsec:threats}

We faced similar validity threats as any other  mapping study.
A first threat to validity is related to the selection of primary sources.
To address this threat we performed our selection based on a repository of over 400 seminal papers that has been collected and classified by leading researchers on the field~\cite{Kientz2013}. 
Furthermore, we extended our selection by specifically searching for terms related to customization in all the major publication outlets.
A second threat is related to the classification scheme. To address this threat, we based our classification along the lines of the taxonomy proposed by Kientz et al.~\cite{Kientz2013} which was complemented with standard terminology of autism research.
A third threat relates to the extraction of data for the classification. For addressing this threat, we carried out our classification using commonly-agreed terms, that were piloted and used by two independent groups.
Subsequently, we held several meetings to discuss our classification results until consensus was reached.

\section{Related Work}
\label{sec:related-work}

The use of technology has had a sizable impact on society and other aspects of life. This impact is also present in educational intervention, seeking the improvement in communication, leisure and curriculum development, changing the daily life of people with autism~\cite{Odom2015}.
This fact could be appreciated in the large number of studies that have been published in the recent years, for example please refer to~\cite{SAGE/GrynszpanWPG14,Hourcade2013,Kagohara2013,Stephenson2015,Bereznak2012,Wainer2011}.
All these works enumerated different technology advances and how they have been applied to support people with autism. However, none of them explores any customization issues.
To the best of our knowledge, customization remains a subject largely unexplored within the realm of autism. There are, however, other projects that aim to improve Human Computer Interaction (HCI) in relation with disabilities in general, but not specific to autism.

The ACCESSIBLE\footnote{http://www.accessible-eu.org} project goal is to research and to develop an Assessment Simulation so that the accessibility of any device designed for web interaction could be easily evaluated. Such evaluation relies on an ontology based on the content of the International Classification of Functioning, Disability and Health (ICF)\footnote{http://www.who.int/classifications/icf/en}, and the Web Content Accessibility Guidelines recommendations\footnote{WCAG1 http://www.w3.org/TR/WCAG10/,  WCAG2 http://www.w3.org/TR/WCAG20/}~\cite{Votis2009}.

The VUMS\footnote{http://vums.iti.gr} cluster is formed by four projects funded by the European Commission: GUIDE , MyUI , VICON  and VERITAS.  The VUMS cluster aims to develop a standard user model considering people with different range of abilities and common data storage format for user profiles.  Although the projects are in the same cluster, each one designs and develops the model representation following different philosophies and according with the specific goals of each project. In GUIDE, the ontology is aimed to a user description as realistic and complete as possible. This approach is very powerful, but introduces a high level of complexity when it comes to runtime instantiation. In the MyUI project the approach is focused on a function based modelling approach rather than on a diagnosis-based approach. Hence, its ontology is based on interaction constraints individual users might have and not on detailed medical impairments and limitations of the individual user. Despite of the different approaches, into the VUMS cluster is defined a VUMS glossary, which is followed by every VUMS project and allows the data interchange~\cite{Moustakas2011}.

The GPII\footnote{http://gpii.net/} project purpose is \emph{"to ensure that everyone who faces accessibility barriers due to disability, literacy, digital literacy, or aging, regardless of economic resources, can access and use the Internet and all its information, communities, and services for education, employment, daily living, civic participation, health, and safety."}.
This is a very ambitious and needed objective, that has received large funding from the governments of US, Canada and the  European Union.   
At the moment, the main product of this work is the result of the project Cloud4All\footnote{http://www.cloud4all.info/} which ended in 2015, and Prosperity4All\footnote{http://www.prosperity4all.eu/} which is still ongoing. 
These projects and other finished projects, make up an overarching system devised in the GPII project whose final results will not be available until project Prosperity4All finishes.  
The Cloud4All project is focused on the instant, ubiquitous auto-personalization of interfaces and materials based on user needs and preferences. Prosperity4All is built on the result of Cloud4All, and is in charge of the development the needed infrastructure for GPII project. With Cloud4All, the data from people, devices and contexts could be defined and represented by an ontology~\cite{Giakoumis2015}. Prosperity4All will access this information to adapt the device interface to the person. Due to the impulse from the governments, and the large efforts performed by many organizations, from different countries, in the development of a whole system, it is expected that many technology developers companies will follow the results of GPII project.

\section{Conclusions and Future Work}
\label{sec:conclusions}

Autism is a neurodevelopmental condition that affects individuals with a large range of combinations of challenges along dimensions such intelligence, social skills, or sensory processing. Hence, any interactive technology for ASD ought to be customizable to fit the particular profile of each individual that uses it. 
In this paper we performed a focused mapping study to assess customization support for this domain along different dimensions, and with emphasis in wearable and natural surfaces technologies. We identified 14 primary sources for our study which allowed us to confirm that even though the importance of customizing technologies for autism is widely accepted and acknowledged, it is still not thoroughly supported.

For future work, as a first step we plan to study the relation of ontology-based representations of accessibility information with the goal of relating them with variability models used for product line engineering. Doing that would help us to apply tools for formal reasoning and development for these technologies. Also, we want to extend the scope of our study to consider recent developments of sensor and wearable technologies~\cite{DBLP:series/synthesis/2016OHara}, machine learning~\cite{BookDeepLearningGoodfellowBC16}, and affective computing~\cite{BookAffectiveComputingCalvoDGK14} could independently and collectively be applied to develop technologies for autism.

\section{Acknowledgments}
This study has been posible thanks to the funding received under the grant agreement 2015-1-ES01-KA201-015946 of the Erasmus+ Program of the European Union and the project Recherche interdisciplinaire sur les syst\`{e}mes logiciels variables sponsored by the \'{E}cole de technologie sup\'{e}rieure, University of Qu\'{e}bec, Canada.

%
\bibliographystyle{abbrv}
\bibliography{biblio}  


\bibliographystyleS{unsrt}
\bibliographyS{primary-sources}

\end{document}